\documentclass[prl,nofootinbib,floats,aps,twocolumn,tightenlines,superscriptaddress,floatfix]{revtex4-1}
\usepackage{graphicx,color}
\usepackage[usenames,dvipsnames]{xcolor}
\usepackage{bbold}
\usepackage{bm}
\usepackage[utf8]{inputenc}
\usepackage{latexsym}
\usepackage{amsmath}
\usepackage{amsthm}
\usepackage{amsfonts}
\usepackage{amssymb}
\usepackage{textcomp}
\usepackage{cancel}
\usepackage{verbatim}
\usepackage[pdftex]{hyperref}
\usepackage{enumerate}
\usepackage[caption=false]{subfig}

\usepackage{amsthm}
 
 \usepackage{dcolumn}

\allowdisplaybreaks

\newcommand{\be}{\begin{equation}}
\newcommand{\ee}{\end{equation}}

\newcommand{\ii}{\mathrm{i}}
\renewcommand{\vec}[1]{\bm #1}

\makeatletter
\newcommand*\nobreakhyphen{\hbox{-}\nobreak\hskip\z@skip}
\makeatother

\makeatletter \def\cleardoublepage{\clearpage\if@twoside \ifodd\c@page\else
   \thispagestyle{empty}
   \newpage
   \if@twocolumn\hbox{}\newpage\fi\fi\fi}
\makeatother

\renewcommand{\d}{\text{d}}

\newcommand{\cnm}[2]{\big[{#1},{#2}\big]}

\theoremstyle{plain}

\theoremstyle{definition}

\newtheorem{e^}{Example}[subsection]

\theoremstyle{remark}

\begin{abstract}
We analyze the implications of the violations of the strong Huygens principle in the transmission of information from the early universe to the current era via massless fields. We show that much more information reaches us through timelike channels (not mediated by real photons) than it is carried by rays of light, which are usually regarded as the only carriers of information.
\end{abstract}

\begin{document}

\title{Violation of the strong Huygen's principle and timelike signals from
the early Universe}
\author{Ana Blasco}
\affiliation{Departamento de F\'{\i}sica Te\'orica II, Universidad Complutense 
de Madrid, 28040 Madrid, Spain}
\author{Luis J. Garay}
\affiliation{Departamento de F\'{\i}sica Te\'orica II, Universidad Complutense 
de Madrid, 28040 Madrid, Spain}
\affiliation{Instituto de Estructura de la Materia (IEM-CSIC), Serrano 121, 28006 Madrid, Spain}
\author{Mercedes Mart\'{\i}n-Benito}
\affiliation{Radboud University Nijmegen, Institute for Mathematics, Astrophysics and Particle Physics, Heyendaalseweg 135, NL-6525 AJ Nijmegen, The Netherlands}
\author{Eduardo Mart\'{\i}n-Mart\'{\i}nez }
\affiliation{Institute for Quantum Computing, University of Waterloo, Waterloo, Ontario, N2L 3G1, Canada}
\affiliation{Department of Applied Mathematics, University of Waterloo, Waterloo, Ontario, N2L 3G1, Canada}
\affiliation{Perimeter Institute for Theoretical Physics, Waterloo, Ontario, N2L 6B9, Canada}

\maketitle

\paragraph{Introduction.---}Quantum field theory in curved spacetimes  provides the natural framework to investigate the quantum nature of matter in the presence of gravity. 
In the context of relativistic quantum information, quantum field entanglement can be used as a powerful physical resource in the analysis of several phenomena such as detection of spacetime curvature or the transmission of information in relativistic settings~\cite{reznik,VerSteeg:2007xs,Collapse2,Comm1,Comm2}.

Quantum entanglement might play an important role in the study of  the early universe (for a review, see \cite{cosmoq,review2}). For example, the phenomena  known as  entanglement `harvesting' and `farming' (i.e. swapping of entanglement from a quantum field to particle detectors \cite{reznik,farming}) is strongly influenced by the cosmological background as proven by Ver Steeg and Menicucci \cite{VerSteeg:2007xs}. 

Another interesting result comes from the consequences in relativistic quantum communication of the violations of the strong Huygens principle  \cite{Comm2}. This principle states that the radiation Green's function  has support only on the light cone \cite{Ellis,McLenaghan,Sonego:1991sq,czapor}. As a consequence,  communication through massless fields is confined to the light cone. This is true in four-dimensional flat spacetime but not in the presence of curvature. These violations have been studied before  in the context of cosmology for classical fields \cite{Faraoni:1991xe, Faraoni:1999us}.  

The violation of the strong Huygens principle implies that there can be a leakage of information towards the inside of the light cone, even for massless quantum fields. When this happens, it is possible to broadcast a message to arbitrarily many receivers with the  energy cost being spent by the receivers of the message \cite{Comm2}. Here we will argue that the violation of the strong Huygens principle has unexpected consequences in cosmological scenarios, 
in particular in the propagation of information from the early universe to the current era.

We will study conformal and minimal couplings of a test massless scalar field in a cosmological background. 
We will show that while the  conformal case does not allow for the leakage of information into the future light cone, the minimal coupling generically allows for information-carrying violations of the strong Huygens principle, in a similar way as it was anticipated in~\cite{Comm2}.
This leads us to the following conclusion: Signals received today  
will generically contain overlapped information about the past from both timelike and light connected events.

Furthermore, the information propagating in the timelike zone decays slower than what one would expect as the spatial distance between sender and receiver increases, as opposed to the information carried by light.  Remarkably, we will see that in a matter-dominated universe, the information propagating in the interior of the light cone does not decay at all with the sender-receiver spatial distance.

\paragraph{Set-up.---} We will consider a spatially flat  and open Friedmann-Robertson-Walker (FRW) spacetime:
\be \label{eq:FRW}
\d s^2=a(\eta)^2(-\d \eta^2+\d r^2+r^2\d\Omega^2).\ee 
Natural units $\hbar = c = 1$ are used throughout. 
This geometry is generated by a perfect fluid with a constant pressure-to-density ratio $p/\rho=w>-1 $, so that the scale factor evolves as $a\propto\eta^{\alpha+{1}/{2}}\propto t^{{(2\alpha+1)}/{(2\alpha+3)}}$, with \mbox{$\alpha=(3-3w)/(6w+2)>-3/2$}. We note that for all $w>-1$, these cosmologies display a Big Bang singularity. Here $t$ is the comoving time $\mathrm{d}t=a(\eta)\text{d}\eta$. 
For computational simplicity, in this background we will consider a test massless scalar field $\Phi$ quantized in the adiabatic vacuum \cite{Birrell}. Notice that although the adiabatic vacuum is in itself an interesting object of study, the choice of the  field's state is not relevant for our results as pointed out below.
We also introduce a couple of comoving observers Alice and Bob. They can perform indirect measurements on the field by locally coupling particle detectors.  

For arbitrary detector trajectories, the interaction between the field and the particle detectors will be described by the  Unruh-DeWitt  model \cite{DeWitt}, which displays all the fundamental features of the light-matter interaction when there is no exchange of orbital angular momentum \cite{Alvaro,Wavepackets}.
The corresponding interaction Hamiltonian (in the interaction picture) for each detector is given by
\be \label{eq:HI}
{H}_{I,\nu}=\lambda_\nu \chi_\nu(t)   \mu_\nu(t)\!  \int \!\text{d}^3 \vec{x} \, a(t)^3F[\bm x-\bm x_{\nu}(t),t]    \Phi[\bm x,\eta(t)],\ee
where $\nu=\{A,B\}$ denotes either Alice's or Bob's detector. 
 $0\le\chi_\nu(t)\le 1$ is the switching function of the detector $\nu$,
$\lambda_\nu$ is its coupling strength, and  $\mu_\nu(t)= |e_\nu\rangle\langle g_\nu| e^{\ii\Omega_{\nu} t} +|g_\nu\rangle\langle e_\nu| e^{-\ii\Omega_{\nu} t}$ is its monopole moment (where $|g_\nu\rangle$ and $|e_\nu\rangle$ are its ground and excited states, and $\Omega_\nu$ is its energy gap). $\bm x_{\nu}(t)$ is the detector's trajectory, which for the  comoving case becomes $\bm x_\nu=$ const.
The field operator $  \Phi$ is evaluated along the worldline of the comoving detectors, which are spatially smeared according to a Gaussian distribution \mbox{$F(\bm x,t)=(\sigma \sqrt{\pi})^{-3}e^{- a(t)^2 \vec{x}^2/\sigma^2 }$},
where $\sigma$ characterizes the constant physical size of the detector. This profile also regularizes the UV divergences that appear in the case of point like detectors \cite{Wavepackets}.

\paragraph{Signalling.---}

In order to study whether Alice and Bob will be able to communicate through the field, we analyze the signalling estimator introduced in \cite{Comm2}, which determines how the excitation probability of $B$ is modulated by the interaction of $A$ with the field. Let \mbox{$|\psi_{0,\nu}\rangle=\alpha_\nu|e_\nu\rangle+\beta_\nu|g_\nu\rangle$} be the initial state of the detector $\nu$. At leading order in time-dependent perturbation theory, this estimator reads $S=\lambda_A\lambda_B S_2+\mathcal O(\lambda_\nu^4)$, where
 \begin{align}\label{signalling}
S_2\!&=\!4\!\int\! \d v\!\!  \int \!\d v' \chi_A(t)\chi_B(t') \text{Re}(\alpha^*_A\beta_A e^{\ii \Omega_A t})F(\bm x-\bm x_A,t)\nonumber\\&\!\!\!\!\!\!\!\!\!\times  \!F(\bm x'-\bm x_B,t') \text{Re}\left(\alpha^*_B\beta_B e^{\ii \Omega_B t'} \cnm{\phi(\bm x,t)}{\phi(\bm x',t')}\right),
\end{align}
and $\text{d}v=a(t)^3\,\text{d}^3\bm x\, \text{d}t$. This expression generalizes the corresponding expression in \cite{Comm2} to our case of smeared detectors,  and as it can be seen in \cite{Comm1}, it is independent of the initial state of the field.
Let us study the form of the field commutator in the cosmological spacetime that we are considering, both for conformally and  minimally coupled massless scalar fields.

For conformal coupling, field modes are given ---except for an overall $1/a(t)$  factor--- by plane waves in conformal time $\eta$. Therefore, the commutator is the same as in Minkowski spacetime, except for overall conformal factors, and vanishes if the events $(\bm x,t)$ and $(\bm x',t')$ are not light connected. Hence, there is no violation of the strong Huygens principle \cite{McLenaghan}: Communication is only possible strictly on the light cone.

In contrast, for minimal coupling, the above commutator does not generically have support only on the light cone. For the cosmological spacetime \eqref{eq:FRW},
\begin{align}\label{com}
\cnm{\phi(\bm x,t)}{\phi(\bm x',t')}&=\ii\frac{\theta( -\Delta\eta)-\theta(\Delta\eta)}{\pi^2 a(t)a(t')R}\\
\nonumber&\times\int_0^\infty \d k\sin(kR) g_{\alpha}\big(\eta(t), \eta(t'), k\big),
\end{align}
where, $R=\|\bm x-\bm x'\|$, $\Delta\eta=\eta(t)-\eta(t')$, and
$g_\alpha$ can be expressed as rational functions of first and second kind Bessel functions  as follows:
\begin{align}
g_{\alpha}(\eta, \eta', k)&=\sqrt{\frac{\eta}{\eta'}}\Big[\mathcal G^{JY}_{\alpha}(\eta,\eta',k)+\mathcal G^{YJ}_{\alpha}(\eta,\eta',k)\Big],\nonumber\\
\mathcal{G}_{\alpha}^{JY}(\eta,\eta',k) &=\frac{J_{\alpha}(k\eta)Y_{\alpha}(k\eta')}{Y_{\alpha}(k\eta')\mathcal L^J_{\alpha}(k\eta')- J_{\alpha}(k\eta')\mathcal L^Y_{\alpha}(k\eta')},\nonumber\\
\mathcal L^{J}_{\alpha}(k\eta)&=J_{\alpha-1}(k\eta)-J_{\alpha+1}(k\eta),
\end{align}
 and $\mathcal{G}^{YJ}_{\alpha}$ and $\mathcal{L}^Y_{\alpha}$ are defined analogously exchanging the Bessel functions  $J_{\alpha}$ and $Y_{\alpha}$.

The case of a cold-matter-dominated universe, for which $\alpha=3/2$, is of particular interest due to its simplicity. In this case the commutator reduces to \cite{Poisson:2011nh}
\begin{align}\label{com-min}
\cnm{\phi(\bm x,t)}{\phi(\bm x',t')}&=\frac{\ii}{4\pi }\Bigg[\frac{\delta (\Delta\eta+R)-\delta(\Delta\eta-R)}{a(t)a(t')R}\nonumber
 \\*&+\frac{\theta (-\Delta\eta- R) -\theta(\Delta\eta-R) }{a(t)a(t')\eta(t)\eta(t')}\Bigg].
\end{align}
We explicitly see the violation of the strong Huygens principle: The commutator gives a non-vanishing contribution to the signalling estimator even when the events $(\bm x,t)$ and $(\bm x',t')$ are timelike separated, due to the \mbox{$\theta$-term}. Let us note that the $\delta$-term is the commutator of the conformally coupled massless scalar field discussed above, and it decays as the comoving distance $R$ grows. Notice that, on the other hand, the contribution of the commutator inside the light cone ($\theta$-term) does not decay as  $R$ increases. This will have important consequences in the transmission of information from $A$ to $B$.

\begin{figure}
\includegraphics[width=0.48\textwidth]{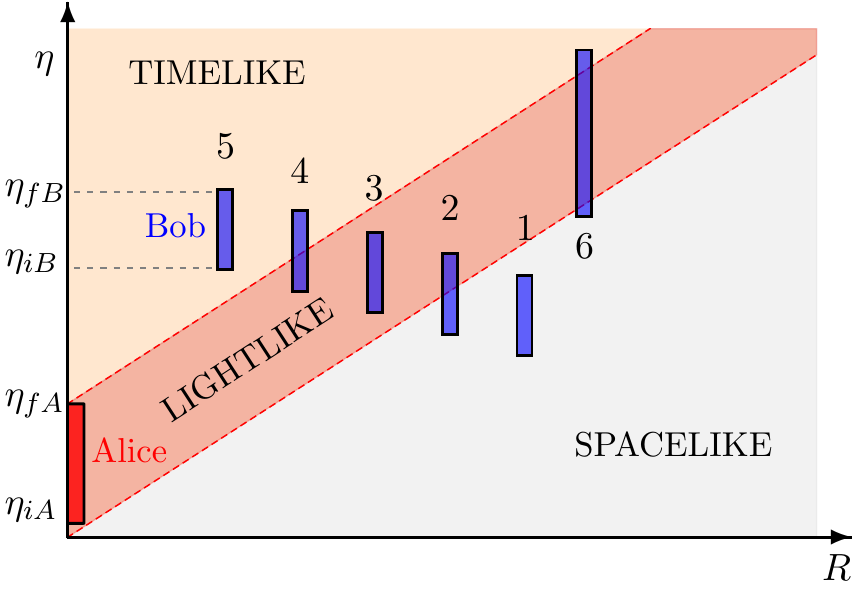}
\caption{Different causal relationships between Alice and Bob's detectors switching periods.   These cases are explicitly specified in Table \ref{tab:table4}. Recall that $\eta_{i\nu}\equiv\eta(T_{i\nu})$, $\eta_{f\nu}\equiv\eta(T_{f\nu})$.}
\label{fig:cases}
\end{figure}

 \paragraph{Pointlike detectors.---}
The probability of excitation of a sharply switched pointlike detector is UV divergent \cite{Louko:2007mu}. However, from the commutator \eqref{com-min} we see that the signalling estimator \eqref{signalling} is UV-safe in the pointlike detector limit ($\sigma\rightarrow0$), even considering sharp switching. Hence, since the pointlike limit is distributionally well behaved, one can take  the abrupt switching function $\chi_\nu(t)=1$ if $t\in [T_{i\nu},T_{f\nu}]$ and zero otherwise. 
The result is finite and given by
\begin{align}\label{S2}
S_{2}&=\frac{1}{\pi}  \text{Re}(\alpha^*_A\beta_A)\text{Im}(\alpha^*_B\beta_B) [S_\delta +S_\theta],
\end{align}
where $S_\delta$ and $S_\theta$ are respectively  the contributions to \eqref{signalling} coming from the Dirac delta and the Heaviside theta terms in \eqref{com-min}. They can be written in terms of polylogarithmic functions  $\text{Li}_s(z)$ for the different cases shown in Fig. \ref{fig:cases} and detailed in Table \ref{tab:table4}. Using the short notation  $\eta_{i\nu}\equiv\eta(T_{i\nu})$, $\eta_{f\nu}\equiv\eta(T_{f\nu})$, their explicit expressions are 
\begin{align}\label{ss}
S_\delta
&=(z_1-z_2)\theta\big(z_1-z_2\big),\\ 
S_\theta&=
\begin{cases}
\ln\left(\frac{\eta_{fA}}
{\eta_{iA}}\right)\ln\left(\frac{\eta_{fB}}
{\eta_{iB}}\right), \;\;\quad\qquad\qquad \mbox{case $5$ (Table \ref{tab:table4})},\\
[L(z_1)-L(z_2)+N(z_1)]\theta(z_1-z_2), \quad \mbox{other cases},
\end{cases} \nonumber
\end{align}
 where  we define 
\begin{align}
& L(z)=\ln\left(\dfrac{R(z-1)}{\eta_{iA}}\right)\ln\left(z\right)+\text{Li}_2\left(1-z\right),\\
& N(z)=\ln\left(\dfrac{R(z-1)}{\eta_{iA}}\right)\ln\left(\dfrac{\eta_{fB}}{Rz}\right),\\
& z_1=\frac{\text{min}\left(\eta_{fA}+R,\eta_{fB}\right)}{R},\quad  z_2=\frac{\text{max}\left(\eta_{iA}+R,\eta_{iB}\right)}{R}.
\end{align}

For simplicity, in  Eqs. \eqref{S2}-\eqref{ss} we have already particularized the study to the case of zero-gap detectors, $\Omega_\nu=0$.  This choice is arbitrary and has no effect on our main results. Moreover, it is not uncommon to find relevant atomic transitions between degenerate (or quasi-degenerate) atomic energy levels, for example, atomic electron spin-flip transitions.

 \begin{table}[b]
\caption{\label{tab:table4}  Cases of causal relationships. See Fig. \ref{fig:cases}.}
\begin{ruledtabular}
\begin{tabular}{cl}
\qquad Case&\qquad\qquad Conditions\\
\hline
  \qquad1&$\eta_{fB}\leq \eta_{iA}+R $\\[0.1 cm]
 \qquad 2&$\eta_{iB}<\eta_{iA}+R <\eta_{fB}\leq\eta_{fA}+R \qquad$\\[0.1 cm]
  \qquad3&$\eta_{iB}\geq \eta_{iA}+R$,$\ \eta_{fB}\leq   \eta_{fA}+R $\\[0.1 cm]
\qquad 4 &$\eta_{fB}> \eta_{fA}+R >\eta_{iB}\geq  \eta_{iA}+R $\\[0.1 cm]
\qquad5& $\eta_{iB}\geq \eta_{fA}+R $ \\[0.1 cm]
 \qquad  6&$\eta_{iB}<\eta_{iA}+R $,$\ \eta_{fB}>\eta_{fA}+R$
  \\
\end{tabular}
\end{ruledtabular}
\end{table}

\paragraph{Channel capacity.---}

\begin{figure}
 \includegraphics[width=0.47\textwidth]{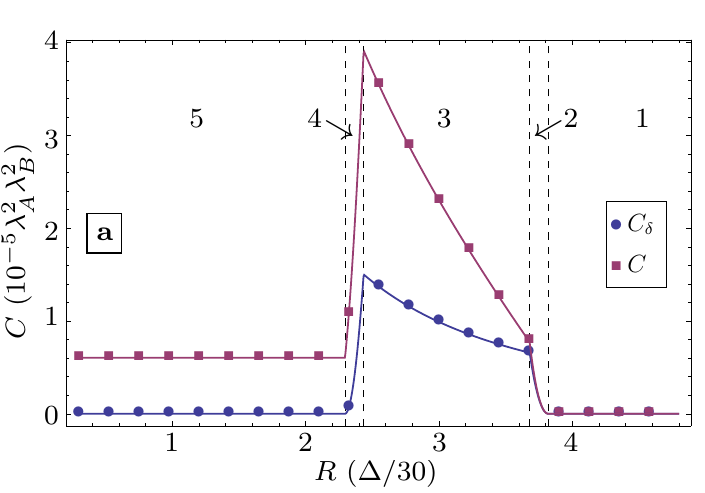}
\includegraphics[width=0.47\textwidth]{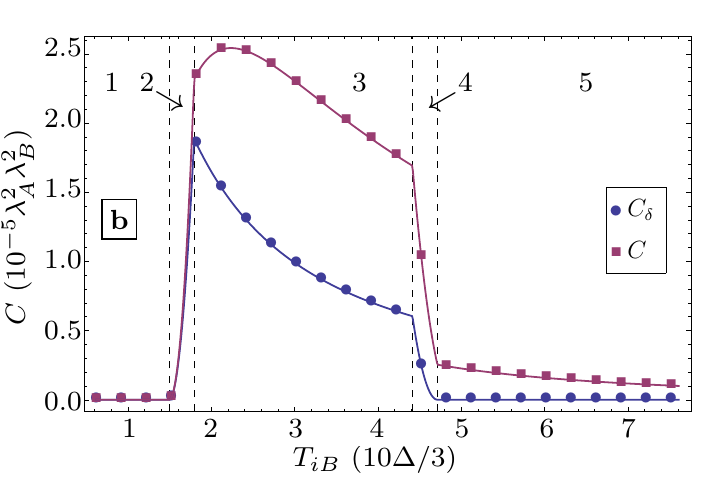}
\caption{ Channel capacity (in bits) and its  $\delta$-term as functions of (a) the spatial separation between Alice and Bob, for \mbox{$T_{fA}-T_{iA}=T_{fB}-T_{iB}=\Delta$},  $T_{iA}= \Delta/30$, and $T_{iB}=10\Delta$, (b)  the temporal separation between Alice and Bob. In (b), we vary $T_{iB}$ while keeping $T_{fA}-T_{iA}=T_{fB}-T_{iB}=\Delta$ constant and we fix $T_{iA}= \Delta/30$ and $R=\Delta/10$. Different regions are labelled according to the case numbers of Fig. \ref{fig:cases} and Table \ref{tab:table4}.
Since both detectors remain switched on during the same amount of proper time, only cases 1 to 5 occur. The violation of strong Huygens can be seen in region 5 (timelike separation).}
\label{fig:Ch_R}
\end{figure}

Let us now compute the capacity of a communication channel between an early Universe observer, Alice, and a late-time observer, Bob. To obtain a lower bound to the capacity, we use a simple communication protocol: Alice encodes ``1'' by coupling her detector $A$ to the field, and ``0'' by not coupling it. Later, Bob switches on his detector $B$ and measures its state. If $B$ is excited, Bob interprets a ``1'', and a ``0'' otherwise.  The capacity of this binary asymmetric channel (i.e., the number of bits per use of the channel that Alice transmits to Bob with this protocol) was proven to be non-zero \cite{Comm2}, no matter the level of noise, and it is given, at leading order, by
\begin{align}\label{capacity}
C\simeq\lambda^2_A\lambda^2_B\frac{2}{\ln 2}\left(\frac{S_2}{4|\alpha_B||\beta_B|}\right)^2 +\mathcal O(\lambda_\nu^6).
\end{align}
Figures \ref{fig:Ch_R}a and   \ref{fig:Ch_R}b  show the behavior of the channel capacity $C$. For comparison, we also display the channel capacity in the conformally coupled case, $C_\delta$. We have selected  initial detector states that, in our case, maximize the channel capacity (i.e. $|\alpha_A|=|\beta_A|=1/\sqrt{2}$, $\arg(\alpha_A)-\arg(\beta_A)=\pi$,  $\arg(\alpha_B)-\arg(\beta_B)=\pi/2$).

Let us first analyze how the ability of Alice to signal Bob depends on their time separation.  
From the $\delta$-term of Eq.~\eqref{com-min}  we see  that the information transmitted by `rays of light' decays with the distance between $A$ and $B$, becoming negligible for long times. This  yields the unsurprising result that the capacity of the light-like communication channel between the early universe and nowadays becomes negligible. In fact the channel capacity decays essentially with the square of the distance between Alice and Bob. Notice that the information carried by `rays of light' constitutes the only contribution to the channel capacity in the conformally coupled case, where strong Huygens principle is fulfilled (see $C_\delta$ in Figs. \ref{fig:Ch_R}a and   \ref{fig:Ch_R}b).

Very remarkably, the $\theta$-term in \eqref{com-min} does not explicitly decay with the distance between $A$ and $B$. Instead, it is inversely proportional to the conformal time between the Big Bang and both $A$ and $B$. Of course, this means that there will be a late-time decay in the channel capacity (see Figs. \ref{fig:Ch_R}a and   \ref{fig:Ch_R}b). However, this decay can be (over)compensated by deploying a number of spacelike separated  $B$  receivers, which fill the interior of  Alice's light cone in a given time slice. This does not entail increasing the information gathered by every $B$ receiver, but instead implies that every $B$ could be regarded as an approximately independent user of the channel that could combine their statistics with the other receivers later on.
Notice that there would be some entanglement harvesting between these spacelike separated $B$ receivers \cite{reznik,VerSteeg:2007xs}, which would in turn correlate their outcomes to some extent. Nevertheless, these harvesting correlations can be made small (e.g. turning down $\lambda_B$ while keeping $\lambda_A\lambda_B$ constant) so that the $B$'s become approximately independent users of the channel. This (over)compensation is possible because the number of receivers that are in timelike contact with $A$ increases as the volume of the light cone (proportional to $ \eta$) while the signalling term decays only logarithmically, as we can  see in  case 5 of $S_\theta$ in Eq. \eqref{ss} as well as in region 5 of Fig. \ref{fig:Ch_R}b.

\paragraph{Conclusions.---} We have studied how the violations of the strong Huygens principle in quantum communication, proposed in  \cite{Comm2}, allow for the transmission of information from the early stages of the Universe to nowadays.

We have focused on a simple lower bound to the capacity of a communication channel between an early Universe observer and a late-time observer who use Unruh-DeWitt particle detectors to transmit and receive information through their local interaction with a massless quantum field. 

We have seen, on very general grounds, that the violation of the strong Huygens principle enables  the transmission of information between events that are timelike separated. This is so even though the receiver cannot receive real quanta from the sender. This is a very general phenomenon in cosmological backgrounds, the most notable exceptions being massless fields conformally coupled or minimally coupled to radiation-dominated universes ($\alpha=1/2$).  The cases of universes dominated by perfect fluids for which the strong Huygens principle is not violated are rare, both for massive and massless fields. We have seen this by explicitly evaluating the massless field commutator for spatially flat homogeneous and isotropic spacetimes, and obtained fully-analytic closed expressions for the channel capacity. 

More importantly, we have shown that the transmission of information via timelike violations of the strong Huygens principle decays more slowly than the information carried by `rays of light'. For the particular case of minimal coupling and a cold-matter-dominated universe, the  channel capacity between timelike separated sender and receiver does not decay at all with their spatial separation. We have also shown that the temporal (logarithmic) decay in the amount of information that can be transmitted through the `Huygens channel' can be compensated by deploying a network of receivers spread over the interior of the future light cone of the sender.

Although we studied the simpler case of a scalar field, the strong Huygens principle is violated for the electromagnetic field $A_\mu$ as well \cite{Noonan2}. Interestingly, the electromagnetic tensor $F_{\mu\nu}$ is conformally invariant, and thus it does not display strong Huygens principle violations \cite{noonan}. However, since the coupling of the charged currents with the electromagnetic field is through $A_\mu$, electromagnetic antennas will see the strong Huygens principle violations in the same fashion as they see e.g. the Aharonov-Bohm effect or Casimir forces. 
Even simple protocols, as the one studied here, show that a considerable amount of information is encoded in the Casimir-like interactions (not mediated by real photons \cite{Comm1,Comm2}) between timelike separated events.

In summary, we conclude that all events that usually generate light-signals  also generate timelike signals not mediated by photon exchange, which may in fact  carry more information than light-like signals. In particular, inflationary phenomena, early universe physics, primordial decouplings, etc., will leave a timelike echo that decays slower than the information carried by light. This might allow us in principle to obtain more information about the early universe than simply observing the electromagnetic radiation.

\paragraph{Acknowledgments.---} The authors would like to give many thanks to Achim Kempf and Jorma Luoko for our very helpful discussions and their valuable insights. The authors would also like to thank Robert Jonsson for our fruitful discussions.  L.J.G. and M.M-B. acknowledge financial support from the  Spanish MICINN/MINECO Project No. FIS2011-30145-C03-02 and its continuation No. FIS2014-54800- C2-2-P. M. M-B. 
M.M-B. also acknowledges financial support from the Netherlands Organization for Scientific Research (Project No. 62001772).

\bibliography{biblio}

\end{document}